\documentclass[conference,10pt,twocolumn]{IEEEtran} 
\usepackage{spconf,amsmath,graphicx}

\usepackage[nolist]{acronym} 
\usepackage{pgfplots}
\usepgfplotslibrary{groupplots}
\usepackage{graphicx}
\usepackage{subcaption}
\pgfplotsset{compat=newest}
\usepackage{hyperref}
\usepackage{amsmath, amsbsy, amssymb}
\usepackage{tikzscale}
\usepackage{color}
\usepackage{epigraph} 
\usepackage{circuitikz}
\usepackage{mathtools}
\usepackage[ruled,boxed,lined,linesnumbered]{algorithm2e}
\usepackage[group-separator={,}]{siunitx}
\usepackage{float}
\usepackage{pstricks}
\usepackage{dblfloatfix}
\usepackage{mwe}
\usepackage{booktabs,multirow}
\usepackage{tabularx}
\usepackage{tikz}
\usetikzlibrary{patterns,arrows}
\usepackage{physics}
\usepackage{verbatim}
\usepackage{subcaption}
\usepackage{wrapfig}

\newcommand{\tredddd}[1]{\textcolor{black}{#1}}

\SetKwRepeat{Do}{do}{while}
\SetKwInOut{Input}{Input}
\SetKwInOut{Output}{Output}

\definecolor{color1}{RGB}{228,26,28}
\definecolor{color2}{RGB}{55,126,184}
\definecolor{color3}{RGB}{77,175,74}
\definecolor{color4}{RGB}{152,78,163}
\definecolor{color5}{RGB}{255,127,0}
\definecolor{color6}{RGB}{200,200,200}

\newcommand{\BfPara}[1]{{\noindent\bf#1.}\xspace}

\usepackage{enumitem}
\newcounter{descriptcount}
\newlist{enumdescript}{description}{1}
\setlist[enumdescript,1]{%
  before={\setcounter{descriptcount}{0}%
          \renewcommand*\thedescriptcount{\arabic{descriptcount}}},
        font={\bfseries\stepcounter{descriptcount}Q \thedescriptcount:}
}

\title{seq2seq-SC: End-to-End Semantic Communication Systems with Pre-Trained Language Model}

\name{
    Ju-Hyung Lee$^*$\thanks{$^*$Both authors contributed equally.} \ 
    Dong-Ho Lee$^*$ \ 
    Eunsoo Sheen \ 
    Thomas Choi  \ 
    \textit{Jay Pujara} \ 
}

\address{University of Southern California \quad\quad\quad
}

\begin{document}

\maketitle

\begin{acronym}
 \acro{AWGN}{additive white Gaussian noise}
 \acro{BER}{bit error rate}
 \acro{BLER}{block error rate}
 \acro{SER}{symbol error rate}
 \acro{DL}{deep learning}
 \acro{GPU}{graphic processing unit}
 \acro{ISI}{inter-symbol interference}
 \acro{LOS}{line-of-sight}
 \acro{MIMO}{multiple-input multiple-output}
 \acro{ML}{machine learning}
 \acro{MLP}{multilayer perceptron}
 \acro{MSE}{mean squared error}
 \acro{NN}{neural network}
 \acro{QAM}{quadrature amplitude modulation}
 \acro{ReLU}{rectified linear unit}
 \acro{SNR}{signal-to-noise ratio}	
 \acro{LLR}{log-likelihood ratio}
 \acro{AE}{autoencoder}
 \acro{ECC}{error correction code}
 \acro{FEC}{forward error correction}
 \acro{LDPC}{low-density parity-check}
 \acro{TX}{transmitter}
 \acro{RX}{receiver}
 \acro{DNN}{deep neural network}
 \acro{CE}{cross entropy}
 \acro{KL}{Kullback-Leibler}
 \acro{wrt}[w.r.t.\@]{with respect to} 
 \acro{i.i.d.}{independent and identically distributed}
 \acro{MMSE}{minimum mean square error}
\end{acronym}

\begin{abstract}
In this work, we propose a realistic semantic network called seq2seq-SC, designed to be compatible with 5G NR and capable of working with generalized text datasets using a pre-trained language model. The goal is to achieve unprecedented communication efficiency by focusing on the meaning of messages in semantic communication. We employ a performance metric called semantic similarity, measured by BLEU for lexical similarity and SBERT for semantic similarity. Our findings demonstrate that seq2seq-SC outperforms previous models in extracting semantically meaningful information while maintaining superior performance. This study paves the way for continued advancements in semantic communication and its prospective incorporation with future wireless systems in 6G networks.
\end{abstract}

%

\section{Introduction} 
\label{sec:intro}

The recent rise of deep learning-based techniques to infer semantics (\textit{i.e.,} the meaning of the message rather than the message itself) from texts, speeches, and videos, as well as the ever-increasing quality of service requirements for extremely data-hungry applications such as extended reality (XR), have motivated the use of semantic communication~\cite{Semantic_Survey2021} for a new generation of wireless systems (6G). Focusing on semantics allows forgoing unnecessary data (\textit{e.g.,} articles in a sentence or background in a portrait photo), which can increase communication efficiency.

While semantic communication may bring unprecedented benefits, many challenges remain to realize it for actual usage. 
First, it must be compatible with existing communication infrastructure; a “link-level” simulation is hence required to verify its realistic end-to-end (E2E) performance. 
Second, the semantic network has to be generalized to work with any dataset rather than a particular dataset. 
Third, since there is no universal performance metric for semantic communication yet, metrics such as semantic similarity must be refined to evaluate performance of the semantic network from the semantic point of view. 
Lastly, since classic communication cannot be completely replaced by semantic communication, the network must be able to deliver information both as-is or modified with high semantic similarity dependent upon the communication scenario.

\BfPara{Contributions}
We revisit questions raised by DeepSC \cite{Semantic_Geoffrey2021} regarding semantic communication:
\begin{enumdescript}
    \item \vspace{0mm} \textit{How do we design the semantic and channel coding jointly?}
    \item \vspace{0mm} 
    \textit{How do we measure semantic error (similarity) between transmitted and received sentences?}

\end{enumdescript}
Our main contributions, which address these questions, are summarized as follows:
\begin{itemize}

\item  \vspace{0mm} We employ E2E link-level simulation compliant to 5G NR (NVIDIA Sionna \cite{hoydis2022sionna}), which contains features like Polar codes. Through such method, we validate semantic network performance in real-world settings and answer \textbf{Q 1}. 

\item  \vspace{0mm} We integrate the pre-trained encoder-decoder transformers with the E2E semantic communication systems, which efficiently extract the semantic (meaningful) information with reduced computation effort, dubbed seq2seq-SC. This network is ``generalized'', meaning it works with all (general) text corpus in comparison to Deep-SC which has limitations dependent upon a particular datasets. 

\item  \vspace{0mm} To answer \textbf{Q 2} and evaluate performance of a semantic network in a \emph{semantic way}, we introduce a metric called semantic similarity. In order to make the network flexible with respect to the communication scenario, the network may either prioritize delivering a message as perfect as possible or on semantic similarity only. 








\end{itemize}



\section{Pre-Trained Model for Language} \label{sec:Background}

Contextualized embeddings from pre-training in a self-supervised manner (\textit{e.g.,} masked language modeling) with transformers~\cite{vaswani2017attention} are extremely effective in providing initial representations that can be refined to attain acceptable performance on numerous downstream tasks.
Recent studies on text semantic communication exploit transformer architectures~\cite{vaswani2017attention} to extract semantics at transmitter and recover the original information at receiver~\cite{Semantic_Geoffrey2021, hu2022one}.
However, such frameworks may have following challenges:
(1) Training an E2E semantic communication pipeline requires a huge computational effort due to the many randomly initialized parameters of semantic encoder/decoder to be trained on;
(2) Difficulty in handling out-of-vocabulary (OOV) since they only use a set of whitespace-separated tokens in the training data.
In this work, we use a pre-trained encoder-decoder transformer~\cite{lewis-etal-2020-bart} to initialize the parameters of semantic encoder/decoder so that the pipeline itself requires little or no computational effort, and use pre-trained tokenizer to effectively handle OOV so that our pipeline can be generalized to any other text.

\section{seq2seq-SC: Semantic Communication Systems with Pre-Trained Model} \label{sec:Body}

\begin{figure}[]
    \centering
    \subfloat[DeepSC \label{fig:DeepSC}]{\includegraphics[width=1\columnwidth]{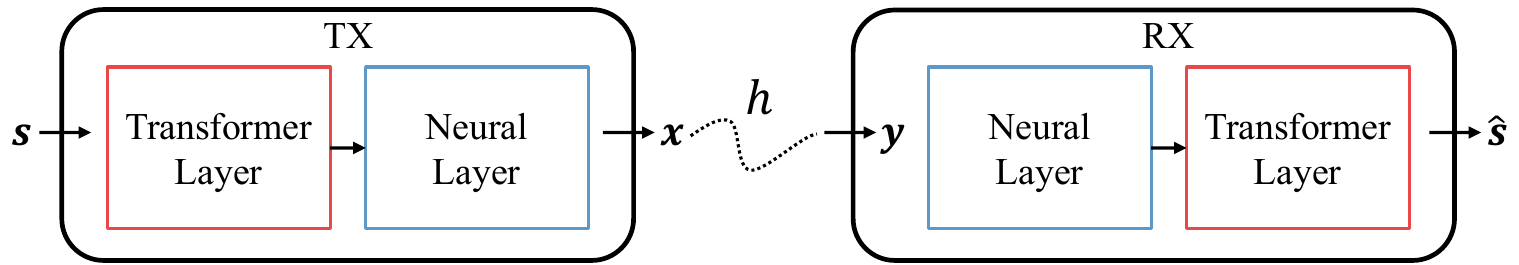}}\\
    \subfloat[seq2seq-SC (Proposed) \label{fig:seq2seq-SC}]{\includegraphics[width=1\columnwidth]{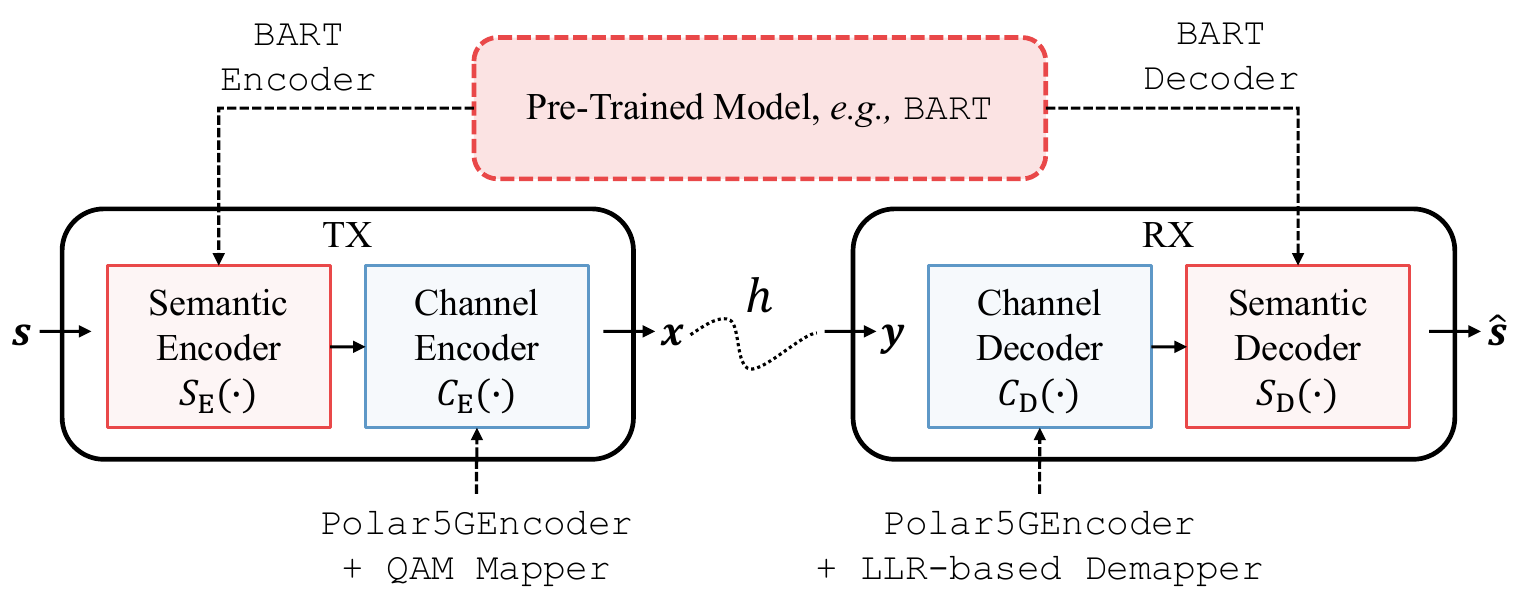}} \\ 
\caption{Architecture of semantic communication systems. \tredddd{Our proposed seq2seq-SC follows the way of link-level simulation, where the channel en/decoder are composed of 5G-NR compliant modules, reflecting the actual symbol (or bit) transmission. In contrast, DeepSC comprises all main modules with neural layers.}}
    \label{fig:system architecture}
\end{figure}

\subsection{Problem Description}
Consider a sentence $\vb* s$ that maps to symbol stream $\vb* x$:
\begin{equation}
	{\vb*{x}} = C_{\mathrm{\mathbf{E}}} \left( S_{\mathrm{\mathbf{E}}} \left( \vb*{s} \right) \right),
\end{equation}
where $C_{\mathrm{\mathbf{E}}}(\cdot)$ and $S_{\mathrm{\mathbf{E}}}(\cdot)$ represent the channel encoder and the semantic encoder, respectively. This symbol stream passes through a physical channel, $h$, with flat fading and noise in the RF front end of a \ac{RX}; which is expressed by the received signal, $\vb* y$: 
\begin{equation}
	{\vb*{y}} = h{\vb*{x}} + {\vb*{n}},
\end{equation}
Here, the encoded signal by \ac{TX} propagates over the Rayleigh fading channel with ${\cal CN}\left( {0,1} \right)$; \ac{RX} receives the attenuated signal with ${\vb* n} \sim {\cal CN}\left( {0,\sigma _n^2} \right)$.
Then, $\vb* y$ is decoded at the \ac{RX} to estimate the sentence $\vb*{\hat{s}}$: 
\begin{equation}
    \vb*{\hat{s}} = S_{\mathrm{\mathbf{D}}} \left( C_{\mathrm{\mathbf{D}}} \left( \vb*{y} \right) \right),
\end{equation}
where $S_{\mathrm{\mathbf{D}}}(\cdot)$ and $C_{\mathrm{\mathbf{D}}}(\cdot)$ represent the semantic decoder and the channel decoder, respectively.

\subsection{Architecture}

The architecture of the semantic communication system is illustrated in Fig. \ref{fig:system architecture}, where the \ac{TX} consists of $C_{\mathrm{\mathbf{E}}}(\cdot)$ and $S_{\mathrm{\mathbf{E}}}(\cdot)$ and \ac{RX} consists of $C_{\mathrm{\mathbf{D}}}(\cdot)$ and $S_{\mathrm{\mathbf{D}}}(\cdot)$.
For the \ac{TX} side, the symbol stream $\vb* s$ is first encoded with a semantic encoder that reduces the size of information by removing information or unnecessary for extracting the necessary information.
Then, it is encoded with a channel encoder, which adds redundancy to quantized source information for reliable detection and correction of bit errors caused by the noisy channel, resulting in $\vb*{x}$.
Inversely, at the \ac{RX} side, a channel decoder first decodes the received signal; and then the semantic decoder decodes the signal and extracts the symbol $\vb*{\hat{s}}$.


\BfPara{Semantic En/Decoder} \quad
For semantic encoder and decoder, we consider a variant of a standard encoder-decoder transformer architecture consisting of two layer stacks in which the encoder is given an input sequence while the decoder generates a new output sequence~\cite{vaswani2017attention}.
Both the encoder and decoder are a stack of $m$ transformer blocks, consisting of a self-attention layer and a fully-connected layer with residual connections, but the self-attention mechanism is different.
The encoder uses a form of fully-visible self-attention mechanism that allows the model to attend to any entry of the input while the decoder uses a form of auto-regressive self-attention which only allows the model to attend to past outputs.
These architectures can be pre-trained on a large scale corpus by corrupting documents and computing the cross entropy loss between the decoder's output and the original document to learn the model generalizable knowledge~\cite{lewis-etal-2020-bart}.
Here, we load such pre-trained checkpoints (\texttt{BART}~\cite{lewis-etal-2020-bart}) and use the weights of encoder and decoder to initialize the weights of $S_{\mathrm{\mathbf{E}}}(\cdot)$ and $S_{\mathrm{\mathbf{D}}}(\cdot)$, respectively.
Also, the pre-trained embedding $\mathcal{E}$ and the pre-trained tokenizer $\mathcal{T}$ are used and shared across $S_{\mathrm{\mathbf{E}}}(\cdot)$ and $S_{\mathrm{\mathbf{D}}}(\cdot)$.
Once the sentence $\vb* s$ is given to $S_{\mathrm{\mathbf{E}}}(\cdot)$, $\mathcal{T}$ tokenizes $\vb* s$ into tokens $\vb* s_{t} = [s_{t_1}, s_{t_2}, ... s_{t_n}]$ and maps each token to embedding $\vb* s_{e} = [s_{e_1}, s_{e_2}, ... s_{e_n}]$ by $\mathcal{E}$.
Then, $S_{\mathrm{\mathbf{E}}}(\cdot)$ encodes $\vb* s$ into hidden states $\vb* r = [r_1, r_2, ... r_n]$ based on $\vb* s_{e}$ and passes it to channel encoder $C_{\mathrm{\mathbf{E}}}(\cdot)$.
Next, hidden states $\vb* r'$, which are recovered from the channel decoder $C_{\mathrm{\mathbf{D}}}(\cdot)$, are given to $S_{\mathrm{\mathbf{D}}}(\cdot)$.
Then, $S_{\mathrm{\mathbf{D}}}(\vb* r')$ defines the conditional distribution $p_{\theta_{S_{\mathrm{\mathbf{D}}}}}\left(\vb*{\hat{s}}_i \mid \vb*{\hat{s}}_{0: i-1}, \vb* r'\right)$ and auto-regressively samples words from the distribution for each index.


\BfPara{Channel En/Decoder} \quad
Polar codes, a form of linear block error correction codes, are one of the channel coding schemes in 5G-NR, where low complexity decoding is available~\cite{3GPP_NR_Polar}.
We consider such practical channel coding modules, \texttt{PolarEncoder} and \texttt{PolarDecoder} as our channel encoder and decoder.
In order to verify our semantic communication systems more practically, other modules (such as modulator and demodulator) are also chosen based on compatibility with 5G-NR.
Details are in Table.~\ref{table_Paramter}.

\begin{table}[!ht]   
  \centering
    \caption{Type (or parameter) for channel en/decoder and network scenario.}
  \resizebox{1.\columnwidth}{!}{\begin{minipage}[t]{.9\columnwidth}
  \centering
  \begin{tabular} {l l}
	\toprule[1pt]
	\textbf{Type (or parameter)} & \textbf{Value} \\
	\cmidrule(lr){1-1} \cmidrule(lr){2-2}
	Channel en/decoder & Polar coding \\ 
	\# of information bits & $512$ \\  
	\# of codeword bits & $1024$ \\  
	Coderate & $0.5$ \\
	Mapper/Demapper constellation & 16-QAM \\ 
	\# of bit per symbol & $4$ \\
	Demapping method & Log-likelihood ratios \\
	Channel & AWGN, Rayleigh fading \\

	\bottomrule[1pt]
\end{tabular}

  \label{table_Paramter}
  \end{minipage}}
\end{table} 

Regarding semantic communication, there are two main goals: (1) the minimization of semantic error, which corresponds to the maximization of semantic capacity (similarity); (2) reduce the number of symbols to be transmitted (\textit{i.e.,} compression).
In this paper, we only focus on minimizing semantic errors by maximizing semantic similarity, which is elaborated on in the following subsection.

\subsection{Evaluation} \label{sec:evaluation}
In traditional communications, the performance of information transmission is evaluated by how accurately bits of 0 or 1 are transmitted (\textit{e.g.,} \ac{BER}) or how well the symbol, which is a set of bits, is transmitted (\textit{e.g.,} \ac{SER}). In contrast, semantic communication focuses on \emph{meaningful information}.
Our view is aligned with the latter angle.
To measure both lexical and semantic similarity, we use BLEU for lexical similarity, and SBERT for semantic similarity.
Note that the higher the SBERT and BLEU scores (in-between $0 \sim 1$), the better. 

\BfPara{BLEU} \quad
BLEU~\cite{papineni-etal-2002-bleu}, originally proposed for machine translation, computes the average of the $n$-gram precision scores by counting the number of matches between $n$-grams of the input and the $n$-grams of the output, in a position independently~\cite{Semantic_Geoffrey2021, hu2022one}.

\BfPara{SBERT} \quad
Even though the lexical similarity between the input and output is low, the semantic similarity can be high.
For example, \textit{``child"} and \textit{``children"} are semantically related, but the BLEU computed lexical similarity is zero.
To compute such semantic similarity, we can represent sentences into embeddings using an embedding model $\boldsymbol{M}$ and compute the cosine similarity between them.

\begin{equation}
\operatorname{match}(\hat{\mathbf{s}}, \mathbf{s})=\frac{\boldsymbol{M}(\mathbf{s}) \cdot \boldsymbol{M}(\hat{\mathbf{s}})^T}{\left\|\boldsymbol{M}(\mathbf{s})\right\|\left\|\boldsymbol{M}(\hat{\mathbf{s}})\right\|}
\end{equation}

Existing semantic communication studies use BERT~\cite{devlin-etal-2019-bert} as $\boldsymbol{M}$ to encode sentences into embeddings and compute the cosine similarity~\cite{Semantic_Geoffrey2021, hu2022one}.
However, the sentence embeddings from such pre-trained models without fine-tuning on semantic textual similarity task poorly capture semantic meaning of sentences due to anisotropic embedding space~\cite{li-etal-2020-sentence}.
Here, we use SBERT~\cite{reimers-gurevych-2019-sentence}, which is fine-tuned on semantic textual similarity tasks, to encode the sentence embedding.


\subsection{Training}
Our framework is trained by cross-entropy loss $\mathcal{L}_{\mathrm{CE}}$ which minimizes the discrepancy between predicted sentence $\vb*{\hat{s}}$ and its original correct input sentence $\vb*{s}$:
\begin{equation}
\begin{split}
    \mathcal{L}_{\mathrm{CE}}(\hat{\mathbf{s}}, \mathbf{s})=
    -\sum_{i=1}\nolimits
    p\left(\mathbf{s}_{t_i}\right) 
    \log 
    \left(p\left(\hat{\mathbf{s}}_{t_i}\right)\right)
    + \\
    \left(1-p\left(\mathbf{s}_{t_i}\right)\right) 
    \log 
    \left(1-p\left(\hat{\mathbf{s}}_{t_i}\right)\right)
\end{split}
\end{equation}
where $p\left(\mathbf{s}_{t_i}\right)$ is the true distribution, where the probability of the correct token for $i$-th index is 1 while other tokens are 0,
and $p\left(\hat{\mathbf{s}}_{t_i}\right)$ is a predicted probability distribution over possible tokens for the $i$-th index.
The framework is trained with
batch size 4 and learning rate 5e-5.



\section{Experiments} \label{sec:exp}


\begin{figure*}[!ht]
\centering
\subfloat[3-gram, \ac{AWGN} \label{a}]{\includegraphics[width=0.5\columnwidth]{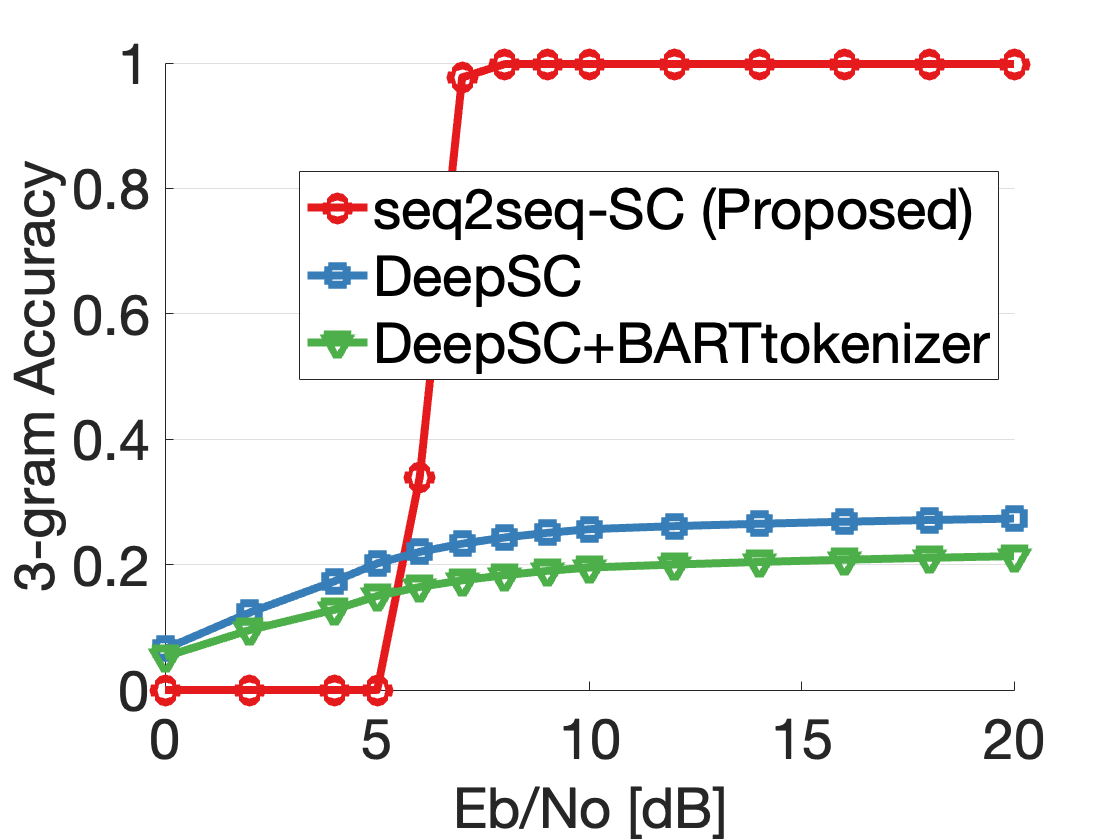}}
\subfloat[4-gram, \ac{AWGN} \label{b}]{\includegraphics[width=0.5\columnwidth]{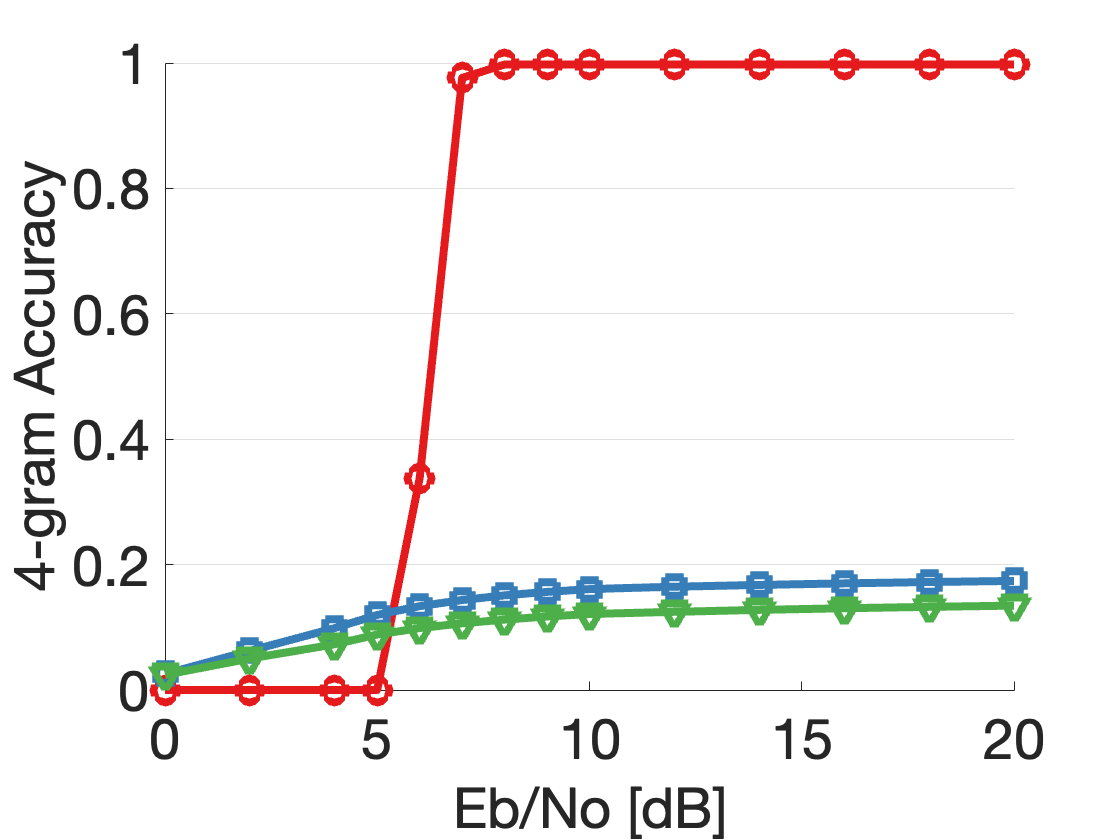}}
\subfloat[3-gram, Rayleigh fading \label{c}]{\includegraphics[width=0.5\columnwidth]{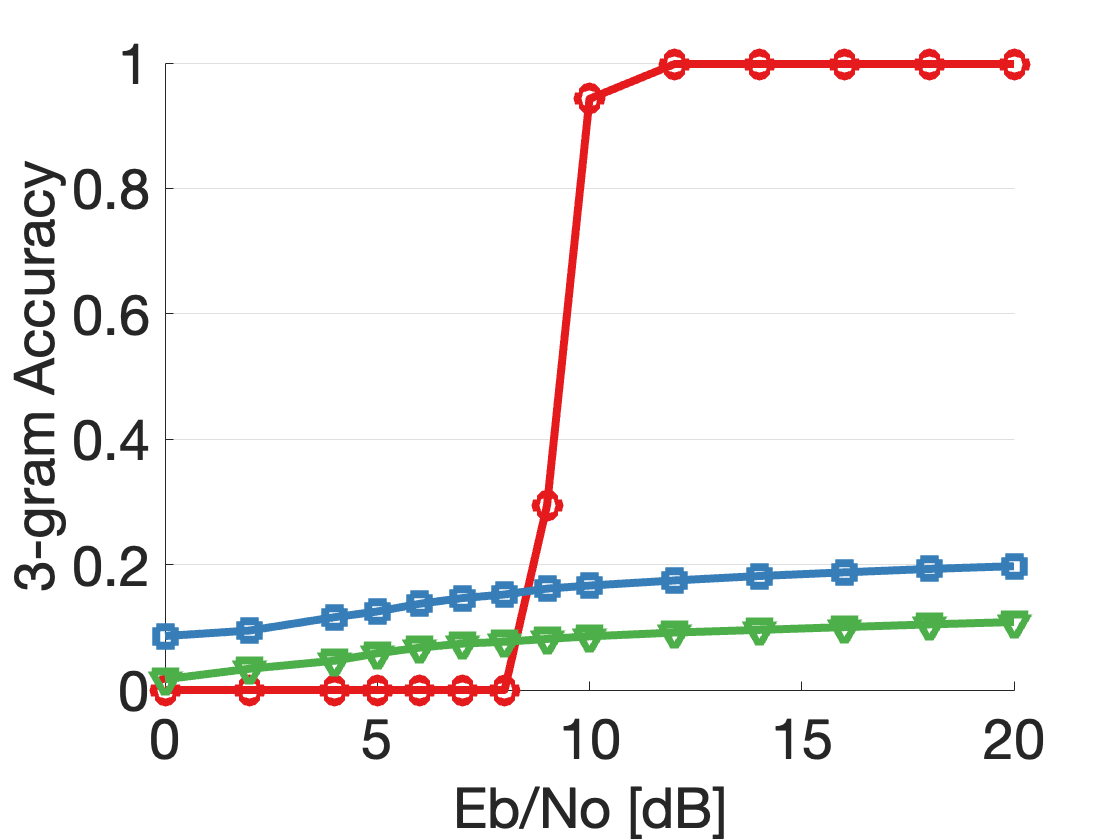}}
\subfloat[4-gram, Rayleigh fading \label{c}]{\includegraphics[width=0.5\columnwidth]{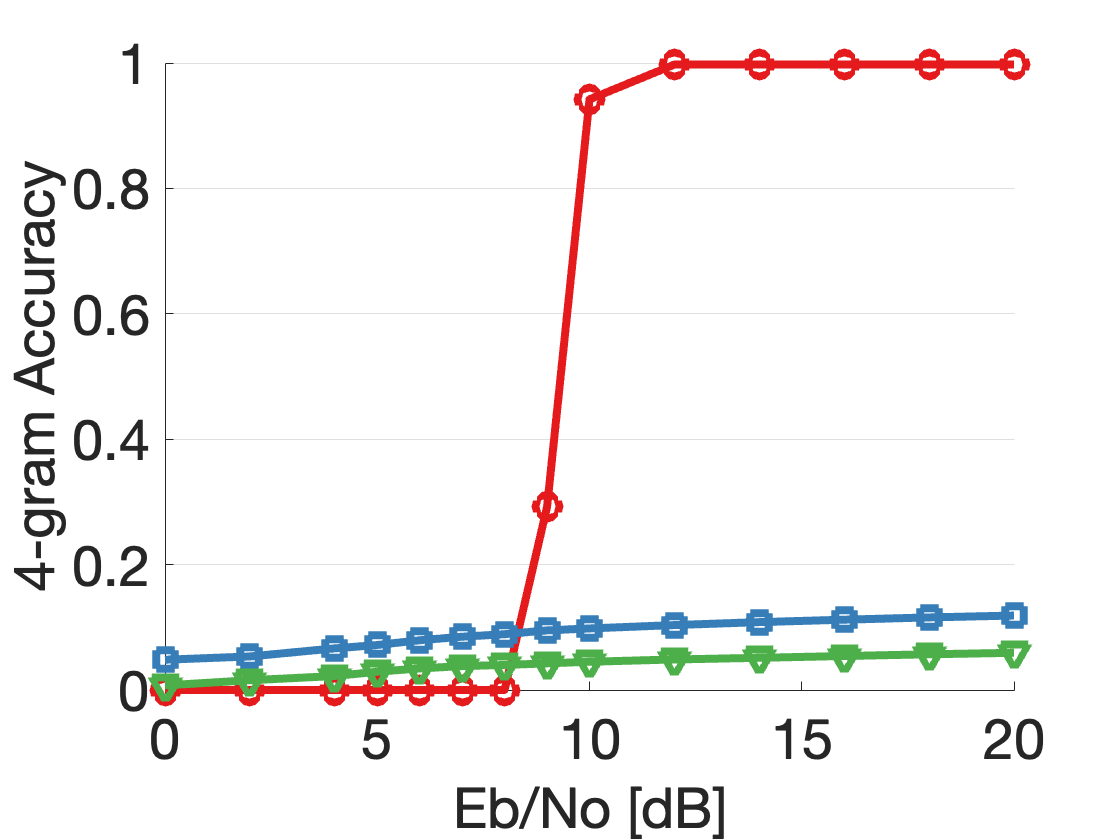}}
\caption{BLEU score over energy per bit to noise spectral density ratio ($E_b/N_o$) [dB] (\ac{AWGN}, Rayleigh fading)}.
\label{fig:comparison}
\end{figure*}

\begin{table*}[!h]
\centering
\resizebox{1.55\columnwidth}{!}{\begin{minipage}[h]{1.3\columnwidth}
\centering
\begin{tabularx}{1\linewidth}{c l l l}
    \toprule[1pt]
    \textbf{Transmit $\vb*{s}$} & An older dog and a younger one playing with a toy. \\
    \midrule[.8pt]
    \textbf{Train} & \textbf{Receive $\hat{\vb*{s}}$} & \textbf{BLEU} & \textbf{SBERT}\\
    \cmidrule(lr){1-1} \cmidrule(lr){2-2} \cmidrule(lr){3-3} \cmidrule(lr){4-4} 
    $\vb*{s} \rightarrow \vb*{s}$  &  An older dog and a younger one playing with a toy. & 1.0 & 1.0 \\
    $\vb*{s} \rightarrow \vb*{s'}$ &  Two dogs are playing with a toy. & 0.210 & 0.820 \\
    \bottomrule[1pt]
\end{tabularx}

\end{minipage}}
\caption{Received output examples of seq2seq-SC trained by different dataset (\textit{i.e.,} $\vb*{s} \rightarrow \vb*{s}$, $\vb*{s} \rightarrow \vb*{s'}$) for $E_b/N_o = 10$\,[dB]. BLEU and SBERT score show the lexical and semantic similarity between the transmitted input $\vb*{s}$ and the received output $\hat{\vb*{s}}$ }
\label{table:exampleSentences}
\end{table*}

\subsection{Datasets}
Following the conventional semantic communication works~\cite{Semantic_Geoffrey2021, hu2022one}, we use the European Parliament dataset~\cite{koehn-2005-europarl}, which consists of $2$M sentences, for model training.
Here, we let the input $\vb*{s}$ and the output $\vb*{\hat{s}}$ be the same for each sentence (\textit{i.e.,} $\vb*{s} \rightarrow \vb*{s}$).
Furthermore, to check whether the input $\vb*{s}$ can be transferred to the receiver in a modified version with the same semantic meaning, we use $270$K pairs of entailment relationship (\textit{e.g., ``A soccer game with multiple males playing.'' $\leftrightarrow$ ``Some men are playing a sport.''}) in natural language inference data~\cite{snli:emnlp2015,N18-1101}.
Here, the output $\vb*{\hat{s}}$ is different from $\vb*{s}$ but the semantic meaning of $\vb*{\hat{s}}$ and $\vb*{s}$ are the same (\textit{i.e.,} $\vb*{s} \rightarrow \vb*{s'}$).
To evaluate the model, we randomly sample 1K sentences from image-caption dataset Flickr~\cite{young-etal-2014-image}, which are not presented in the training data, to check the generalizability and superiority of our framework.

\subsection{Baselines}
We compare our model with the following models:
(1) \textbf{DeepSC}~\cite{Semantic_Geoffrey2021} consists of semantic en/decoder including multiple transformer en/decoder layers, respectively, while each channel en/decoder uses dense layers; that is, \ac{DNN}-based E2E physical layer communication systems.
Framework tokenizes sentences in the training data and creates a set of tokens that can assign an embedding to each token in the training data;
(2) \textbf{DeepSC+BARTtokenizer} is designed for a more fair comparison with our framework; it replaces the tokenizer (1) with a pre-trained BART tokenizer~\cite{lewis-etal-2020-bart} to process tokens that are not in the training data;
(3) \textbf{seq2seq-SC} is our main model consisting of semantic en/decoder initialized with pre-trained model checkpoint (\texttt{BART-base}), pre-trained BART tokenizer~\cite{lewis-etal-2020-bart}, and the channel en/decoder with Polar coding.
Other modules in \ac{TX} and \ac{RX} (\textit{e.g.,} mapper/demapper) are NR-5G compatible and configured based on the link level simulator \texttt{NVIDIA Sionna} \cite{hoydis2022sionna}.

\subsection{Experimental Results} \label{sec:Num}



\BfPara{Comparison Study: seq2seq-SC vs DeepSC}
Fig. \ref{fig:comparison} shows the BLEU scores of DeepSC  and our proposed seq2seq-SC for 1K sentences sampled from Flickr~\cite{young-etal-2014-image} not used for training.
DeepSC achieves only about 10 to 20\% 3,4-gram accuracy in the unseen corpus
while our proposed seq2seq-SC outperforms it, achieving a near-perfect lexical similarity for $E_b/N_o \geq6$\,[dB]. 
This highlights the generalizability and superiority of our proposed framework.

\tredddd{It is worth noting that, for $E_b/N_o < 6$\,[dB], however, the accuracy drops remarkably, as the channel begins to cause an error to the input of the semantic decoder even after the error correction in the channel decoder.
Here, the polar coding, which is our considered \ac{FEC} method in the channel en/decoder, achieves \ac{BER}~$>10^{-3}$ for $E_b/N_o \simeq 5$\,[dB].
The out/input of semantic en/decoder is a tensor whose elements are FP32 single-precision floating points of 32 bits. 
The tensor, the token the semantic en/decoder exchange with, is particularly vulnerable to bit-wise errors (\textit{e.g.,} bit flip); for instance, $1.0$ can become $\infty$ even with a single flip of the second bit; that explains such a drastic accuracy drop.
}
In this experiment, we utilized $\tanh(\cdot)$ function to map the elements of the tensor into $[-1, 1]$, but the accuracy in a low-\ac{SNR} channel can be improved by introducing a better method to handle bit flip in tensors, and it is the subject of our future work.

\begin{table}[!t]
\centering
\resizebox{.95\columnwidth}{!}{\begin{minipage}[h]{.8\columnwidth}
\centering
    
    

\begin{tabularx}{1\linewidth}{ccc}
    \toprule[1pt]
    \multirow{2}{*}{\textbf{Train}} & \textbf{Lexical Similarity} & \textbf{Semantic Similarity} \\
    \cmidrule(lr){2-2} \cmidrule(lr){3-3} 
    & BLEU~\cite{papineni-etal-2002-bleu} & SBERT~\cite{reimers-gurevych-2019-sentence}  \\
    \midrule
    $\vb*{s} \rightarrow \vb*{s}$ & 0.993  & 0.999 \\
    $\vb*{s} \rightarrow \vb*{s'}$ & 0.173 & 0.764  \\
    \bottomrule[1pt]
\end{tabularx}

    


\end{minipage}}
\caption{Lexical and semantic similarities of seq2seq-SC, trained by different training data, $E_b/N_o = 10$\,[dB].}
\label{semanticSimilarity}
\end{table}

\BfPara{Semantic Similarity}
As aforementioned, the interoperability of delivering information both as-is or modified with high semantic similarity, dependent upon the communication scenario, is important. 
Table~\ref{semanticSimilarity} corroborates the interoperability of the proposed seq2seq-SC.
Since there is no universal performance metric for semantic communication yet, here, the semantic similarity is evaluated by several metrics from different semantic points of view.
When the framework is trained to output the same sequence as the input (\textit{i.e.}, $\vb*{s} \rightarrow \vb*{s}$), both the lexical and semantic similarities show a near-perfect score for $E_b/N_o = 10$\,[dB].
\tredddd{However, when the framework is trained to output a sentence that is semantically similar to the input (\textit{i.e.}, $\vb*{s} \rightarrow \vb*{s'}$), the lexical similarity drops significantly while the semantic similarity remains similar, meaning it is a lexically different sentence but is semantically alike to the original sentence.
It shows that our proposed system empowered by pre-trained \texttt{BART} model successfully decodes the original information and extracts semantic (meaningful) information upon their scenario, achieving high lexical and semantic similarities, as demonstrated in Table~\ref{table:exampleSentences}. 
Such interoperability underlines the potential of refined pre-trained models (weights) that could be utilized as a new type of source compression technique, which can be discussed in our future work.}

\section{Conclusions} \label{sec:conclusion}
In this paper, we have shown that seq2seq-SC, which uses a pre-trained model, not only helps with computation time during training, but also with generalization of adapting to new texts unforeseen during training. This network proved its superiority over Deep-SC in terms of semantic similarity, through link-level simulations which closely resemble actual 5G systems. While we have focused on text datasets, such pre-trained model in the future can be expanded to other datasets including speeches, videos, etc., where benefits of semantic communication in terms of communication efficiency becomes even more substantial. 


\vfill\pagebreak

\bibliographystyle{IEEEbib}

\bibliography{strings,refs}

\end{document}